# ON POSSIBLE CAUSES OF DIVERGENCIES IN EXPERIMENTAL VALUES OF GRAVITATIONAL CONSTANT

A.L. Dmitriev

St-Petersburg University of Information Technologies, Mechanics and Optics
49, Kronverksky Prospect, St-Petersburg, 197101, Russia

*It is shown that discrepancies in the experimental values of the gravitational constant might be caused by the temperature dependence of the gravitational force and inequality of the absolute temperatures of sample masses used in various gravitational experiments.*

It is known that some divergences in the absolute values of the gravitational constant obtained by various authors in various experiments considerably exceed the accuracy of measurements [1-3]. Differencies of the experimental average values of the gravitational constant $G$ reaches the level $10^{-4} - 10^{-3}$ as compared to value ($G = 6.6742 \cdot 10^{-11} m^3 kg^{-1} s^{-2}$) recommended by CODATA, and it is usually explained by the effects of hypothetic factors: effect of the Earth's magnetic field, time fluctuations of $G$ value, dependence of $G$ on direction in space, and some others. The temperature dependence of gravity force examined in [4] explains the divergences in the experimental values of $G$ by differences in the absolute temperatures of sample masses used in those experiments. According to [4,5], the temperature dependence of the gravitational constant $G$, in the first approximation, might be presented as

$$G = G_0 (1 - a_1 \sqrt{T_1})(1 - a_2 \sqrt{T_2}), \qquad (1)$$

where $G_0$ – the constant, $a_1$ and $a_2$ – temperature factors depending on the physical properties (density and elasticity) of interacting masses, $T_1$ and $T_2$ – their absolute temperatures exceeding Debye temperature. For heavy and viscous metals (lead, copper, brass) the experimental values of factors $a_{1,2}$ are approximately in the range of $(1.5 - 2.5) \cdot 10^{-4} K^{-1/2}$, and for light and elastic metals (duralumin, titanium) - $(4.0 - 9.5) \cdot 10^{-4} K^{-1/2}$ [4]. If the sample masses have been made of one and the same metal ($a_1 = a_2 = a$) and their average temperatures are also equal ($T_1 = T_2 = T$), the relative temperature shift of the gravitational constant $G$ value being measured is equal to

$$\frac{\Delta G}{G} = \frac{a \Delta T}{\sqrt{T}}, \qquad (2)$$

where $\Delta T$ – temperature change of both masses.

For example, at $a = 5 \cdot 10^{-4} K^{-1/2}$, $T = 293K$ and $\Delta T = 5K$ the relative temperature shift of the gravitational constant $G$ values being measured is equal to $5 \cdot 10^{-4}$. For the sample masses made of light and elastic materials (quartz glass, crystals) the value of the temperature factor might be close to $10^{-3} K^{-1/2}$, then the corresponding relative temperature shift $\Delta G / G$ is in the range of $10^{-4} - 10^{-3}$.

In materials of the papers devoted to measurements of the gravitational constant $G$ values only the stability of the interacting masses temperatures is usually shown but not its absolute value [2,6]. The variance of the average absolute values of the sample masses temperatures in the gravitational experiments carried out in different laboratories practically can not be avoided and might reach a few degrees. Other physical conditions of experiments on determination of the absolute value of $G$ might also essentially differ – different materials, dimensions and form, temperatures of the sample bodies, etc. The numerical estimates given above show that comparatively small (units of degrees) differences in the absolute temperatures of sample masses in different experiments might be one of the imported causes of the experimental average values scattering of $G$ which has been observed for a long time.

It should be especially noted the temperature dependence of the gravitational force [4,5] does not contradict to the known facts of the classical mechanics. Such dependence is of a great importance for the present problems of the high-precision metrology of mass, gravimetry, as well as the astrophysics. Experimental investigations of the body temperature effects on the force of the gravitational interaction of bodies are of current interest and expedient.